\documentclass[useAMS,usenatbib]{mn2e}
\usepackage{graphicx}
\def\rmscr#1{{\hbox{\rm \scriptsize #1}}}
\def\rmmat#1{{\hbox{\rm #1}}}
\def\d{\rmmat{d}}

\begin{document}
\title[White-Dwarf Kicks II.]{Constraining white-dwarf kicks in globular
  clusters : II.~Observational Significance}

\author[J. S. Heyl]{Jeremy S. Heyl$^{1}$\\
$^{1}$Department of Physics and Astronomy, University of British Columbia, Vancouver, British Columbia, Canada, V6T 1Z1 \\
Email: heyl@phas.ubc.ca; Canada Research Chair}

\date{\today}

\pagerange{\pageref{firstpage}--\pageref{lastpage}} \pubyear{2007}

\maketitle

\label{firstpage}

\begin{abstract}
  If the winds of an asymptotic-giant-branch stars are sufficiently
  strong are slightly asymmetric, they can alter the star's trajectory
  through a globular cluster; therefore, if these winds are
  asymmetric, one would expect young white dwarfs to be less radially
  concentrated than either their progenitors or older white dwarfs in
  globular clusters.  This latter effect has recently been observed.
  Additionally the young white dwarfs should have larger typical
  velocities than their progenitors.  After phase mixing this latter
  effect is vastly diminished relative to the changes in the spatial
  distribution of young white dwarfs with kicks, so it is more
  difficult to detect than the change in the spatial distribution.
  The most powerful kinematic signature is the change in the
  eccentricity of the orbits that is revealed through the distribution
  of the position angles of proper motion.
\end{abstract}

\section{Introduction}

\citet{1998A&A...333..603S} argued that their observed rotation rates
of white dwarfs may be evidence that they receive mild kicks from by
asymmetric winds toward the end of their time on the asymptotic giant
branch (AGB) \citep{1993ApJ...413..641V}.  \citet{2003ApJ...595L..53F}
argued that these mild kicks could account for a putative dearth of
white dwarfs in open clusters
\citep[e.g.][]{1977A&A....59..411W,2001AJ....122.3239K}.  The dynamics
of open clusters provide limited evidence because one expects the
white dwarfs simply to leave the cluster.  One can probe the dynamics
of AGB winds directly through observations of masers, but such
attempts are dogged by the variability of the star itself which makes
it difficult to constrain any relative motion between the centre of
mass of the wind and that of the star.  The globular clusters provide
a unique probe of white dwarf kicks because both the magnitude of the
expected kick and the velocity dispersion of the giants within the
cluster are on the order of kilometers per second --- one would expect
a significant signal.  On the other hand the escape velocity of the
cluster may be several times larger so most of the white dwarfs remain
in the cluster to measure after the kick.  The expected signature of
white dwarf kicks has been observed in M4 and NGC~6397
\citep{2007Davis}.

This letter will examine how size of the kick, the initial
distribution of stars and the size of the observational sample affects
the expected statistical significance that white dwarf kicks have on
observations of globular clusters through Monte Carlo calculations
of the phase-space distribution function of stars in the cluster.

\section{Calculations}

\citet{Heyl07kickgc} gives the details of the numerical calculations,
but a quick summary is useful.  Clusters of stars can typically be
modelled with a lowered isothermal profile (or King model)
\citep{1963MNRAS.125..127M,1966AJ.....71...64K}
\begin{equation}
f = \frac{\d N}{\d^3 x \d^3 v} = 
\left \{ 
\begin{array}{ll}
\rho_1(2\pi\sigma^2)^{-3/2} \left ( e^{\epsilon/\sigma^2}- 1 \right )
&  \rmmat{~if~} \epsilon>0 \\
0 & \rmmat{~if~} \epsilon\leq 0 
\end{array}
\right .
\label{eq:1}
\end{equation}
where $\epsilon = \Psi - \frac{1}{2} v^2$, $\Psi$ is the gravitational
potential, $\sigma$ is the characteristic velocity dispersion of the
cluster as a whole and $\rho_1$ is a characteristic density.

With time the kinetic energy within the cluster approaches
equipartition between the various stars such that $m_i \sigma_i^2 =
m_j \sigma_j^2$ \citep{Spit87}.  The progenitors of young white dwarfs
will be the most massive main-sequence stars in a cluster at the time,
so they will typically have $\sigma_\rmscr{TO}<\sigma$, where
$\sigma$ is the mean velocity dispersion of the
cluster; furthermore, because these progenitors will only have a small
fraction of the mass of the cluster, they can be considered as
massless tracers; their phase space density will be given by
Eq.~(\ref{eq:1}) with $\sigma \rightarrow \sigma_\rmscr{TO}$.

During their time on the AGB, the stars may lose a large fraction of
their mass suddenly and asymmetrically
\citep{1993ApJ...413..641V,1998A&A...333..603S,2003ApJ...595L..53F}
and receive an impulsive kick.  Because the distribution no longer
depends on constants of the motion alone, the distribution function
itself will depend on time, so a Monte Carlo realization of the kicked
distribution function is helpful to make further progress.
Specifically, for a variety of values of $\Psi(0)$,
$\sigma_\rmscr{TO}$ and $\sigma_k$, one hundred thousand stars are
drawn from Eq.~(\ref{eq:1}) and given a three-dimensional velocity
kick drawn from a Gaussian of width of $\sigma_k$.  Finally, each star
is evolved forward along its orbit to a random phase.

The result of the Monte Carlo realization is a list of positions and
velocities just before the kick, just after the kick, and at a random
time later.  For each star these positions and velocities are
projected along a random direction and into the plane perpendicular to
that direction, yielding a distribution of velocities and positions
parallel and perpendicular to the line of sight to the globular
cluster.  In principle all three components of the velocity and
position are measurable.  A useful measurement of the distance of a
faint white dwarf along the line of sight requires micro-arcsecond
astrometry; measurements of the velocity along the line of sight
require multiplexing spectroscopy again to faint magnitudes.  Although
these may be feasible in the future, the focus will be on the
quantities perpendicular to the line of sight.  Finally, because the
globular cluster is spherically symmetric the position angle of a star
relative to the centre of the cluster is uniformly distributed.
However, the position angle of the proper motion relative to the
projected direction toward the centre of the cluster does carry useful
information about the properties of the stellar orbits.
 
\section{Results}

This leaves five quantities of interest: the distance and speed of a
star relative the centre of the cluster in the plane of the sky, the
angle between the proper motion of the star and the direction toward
the centre of the cluster and of secondary interest, the distance and
speed of a star relative the centre of the cluster along the line of
sight.  Furthermore, because the line of sight to the observer is
arbitrary the distributions of these latter two quantities are simply
related to the distributions of the astrometric quantities so the
results will focus on the three astrometric measurements.

\subsection{Distributions}

\begin{figure}
\includegraphics[width=3.4in]{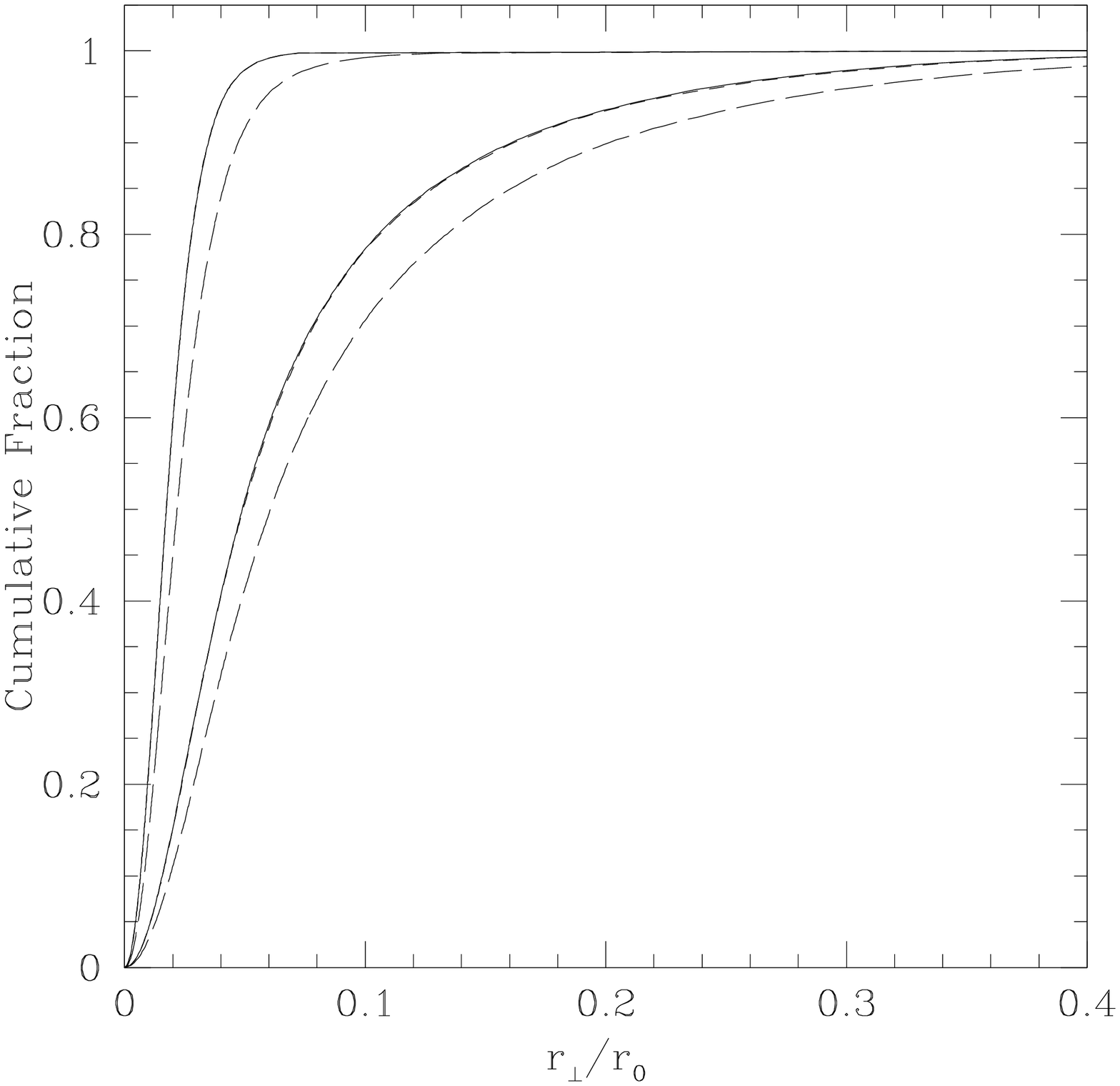}
\includegraphics[width=3.4in]{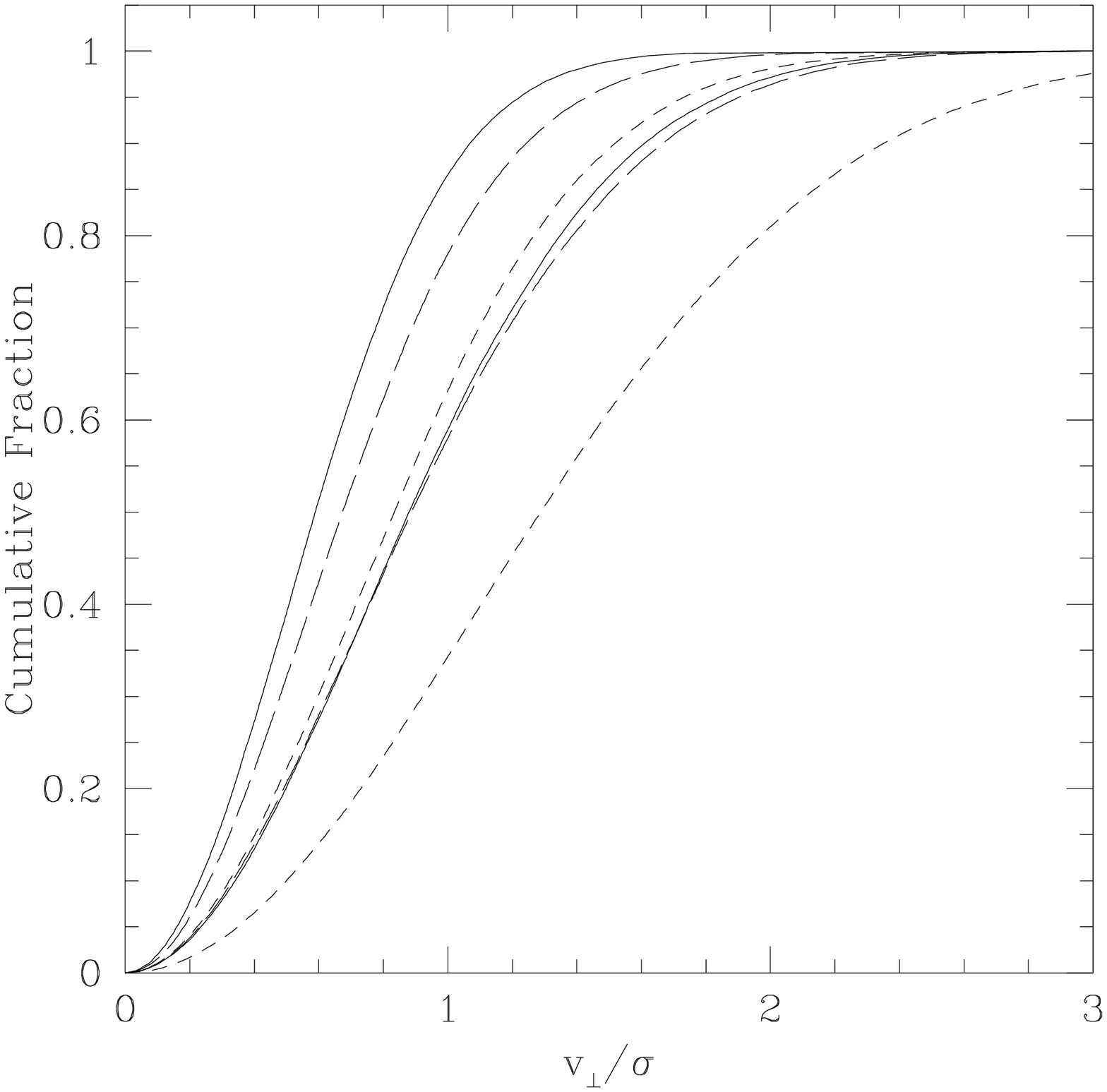}
\caption{The cumulative position and velocity distributions in the
  plane of the sky for
  $\sigma_\rmscr{TO}=\sigma_k=0.5\sigma$ (left curves) 
  and $\sigma_\rmscr{TO}=\sigma_k=0.8\sigma$ (right curves) with $\Psi(0)=6\sigma^2$.
  The solid curves follow the initial distribution, and the
  short-dashed curves follow the distribution immediately after the
  kick (these two coincide for the radial distribution).  The
  long-dashed curves follow the distribution after phase mixing.}
\label{fig:pos_vel}
\end{figure}

The kicks that the white dwarfs receive as they form can dramatically
affect their phase-space distribution.  The upper panel of
Fig.~\ref{fig:pos_vel} shows that a kick of the same magnitude as the
velocity dispersion of the population significantly puffs up the
projected radial distribution of the young white dwarfs.
Additionally, the effect is similar in magnitude regardless of how
tightly bound the progenitors are.  This agrees with the results of 
\citet{Heyl07kickgc} for the three-dimensional radial distribution.

The more subtle effect is the change in the distribution of proper
motions, both in magnitude and direction.  The lower panel of
Fig.~\ref{fig:pos_vel} shows that the change in the distributions of
the magnitude proper motions is smaller especially for the loosely
bound progenitors ($\sigma_\rmscr{TO}=0.8\sigma$).  Immediately after
the kick the velocity distributions of the young white dwarfs spread
out dramatically.  However, after averaging over the phase of the new
orbits, the effect is diminished.  Although this may initially be
surprising, it results from the negative specific heat of
gravitationally bound systems.  For example, if the orbit were
initially circular and Keplerian, the kick would serve to lift the
apocentre of the orbit; therefore, on average the velocity along the
orbit would decrease.  Of course, the orbits are not initially
circular and thew potential is not Keplerian so the net result is to
diminish the effect (and not reverse it).  The reduction is more
dramatic for the loosely bound stars because a modest increase in the
velocity can greatly increase the size of the orbit and possibly
unbind the star from the cluster; therefore, the stars that remain
bound to the cluster will necessarily have lower velocities --- the
loosely bound stars are more strongly affected by evaporative cooling.

\begin{figure}
\includegraphics[width=3.4in]{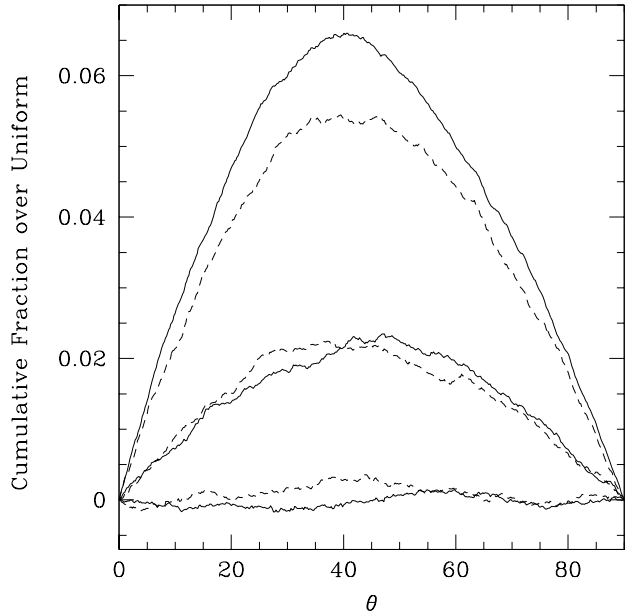}
\caption{The cumulative distribution over uniform of the angle between
  the proper motion and the projected direction toward the centre of
  the cluster for $\sigma_\rmscr{TO}=0.5\sigma$ (solid curves) and
  $0.8\sigma$ (dashed curves) with $\Psi(0)=6\sigma^2$ after phase
  mixing.  From bottom to top the curves trace $\sigma_k=0.5, 1$ and
  1.5 $\sigma_\rmscr{TO}$.  The distributions before the kick and
  before phase mixing are consistent with a uniform distribution.  All
  the distributions for $\sigma_k=0$ are also uniform.  In all cases a
  uniform distribution has been subtracted from the cumulative
  distribution.}
\label{fig:angle}
\end{figure}
The orbits of the stars change in a second way.  The orbits after the
kick and orbit averaging are typically more eccentric that before.
The distribution of velocities according to Eq.~(\ref{eq:1}) is
spherically symmetric.  Furthermore, the kicks are distributed
symmetrically; therefore, the distribution of the angle between the
proper motions and the projection direction to the centre of the
cluster is uniform.  However, after the kicked stars have mixed in
phase, it is more likely to find stars moving within forty-five
degrees of the radial direction than within forty-five degrees of the
tangential direction -- the orbits are typically more eccentric than
before.  Fig.~\ref{fig:angle} shows the cumulative distribution of
this angle ($\theta$) for various sizes of kick with a uniform
distribution subtracted away.  First, the effect weakens with the size
of the kick and vanishes for no kick.  Second, for large kicks the
effect is weaker for the more loosely bound stars.  Again this is
because the loosely bound stars are more likely to get eccentricities
greater than unity and leave the system.

\subsection{Detectability}

Although these differences in the distributions are apparent when many
thousands of stars are use as probes, in reality one typically has
only several tens of stars to characterise the distribution of young
white dwarfs (the kicked stars) and possibly a factor of ten more
stars to estimate the distribution of the progenitors.  To estimate
the typical sensitivity of measurements to the effects of white dwarf
kicks, smaller samples were generated with various numbers of kicked
stars and a factor of ten more progenitors for comparison.  A total
of one thousand samples were generated in each group with
$\sigma_\rmscr{TO}=0.5\sigma$ and $0.8\sigma$ and
$\sigma_k=\sigma_\rmscr{TO}$.  For each pair of samples the likelihood
that they were drawn from the same underlying distribution was
calculated using the Kolmogorov-Smirnov test and the Wilcoxon rank-sum
test.  the final result is the median likelihood over the one thousand
realisations. For the sake of clarity and definiteness observational
errors have not been included in the samples generated -- such errors
would reduce the significance of the detections.
\begin{figure}
\includegraphics[width=3.4in]{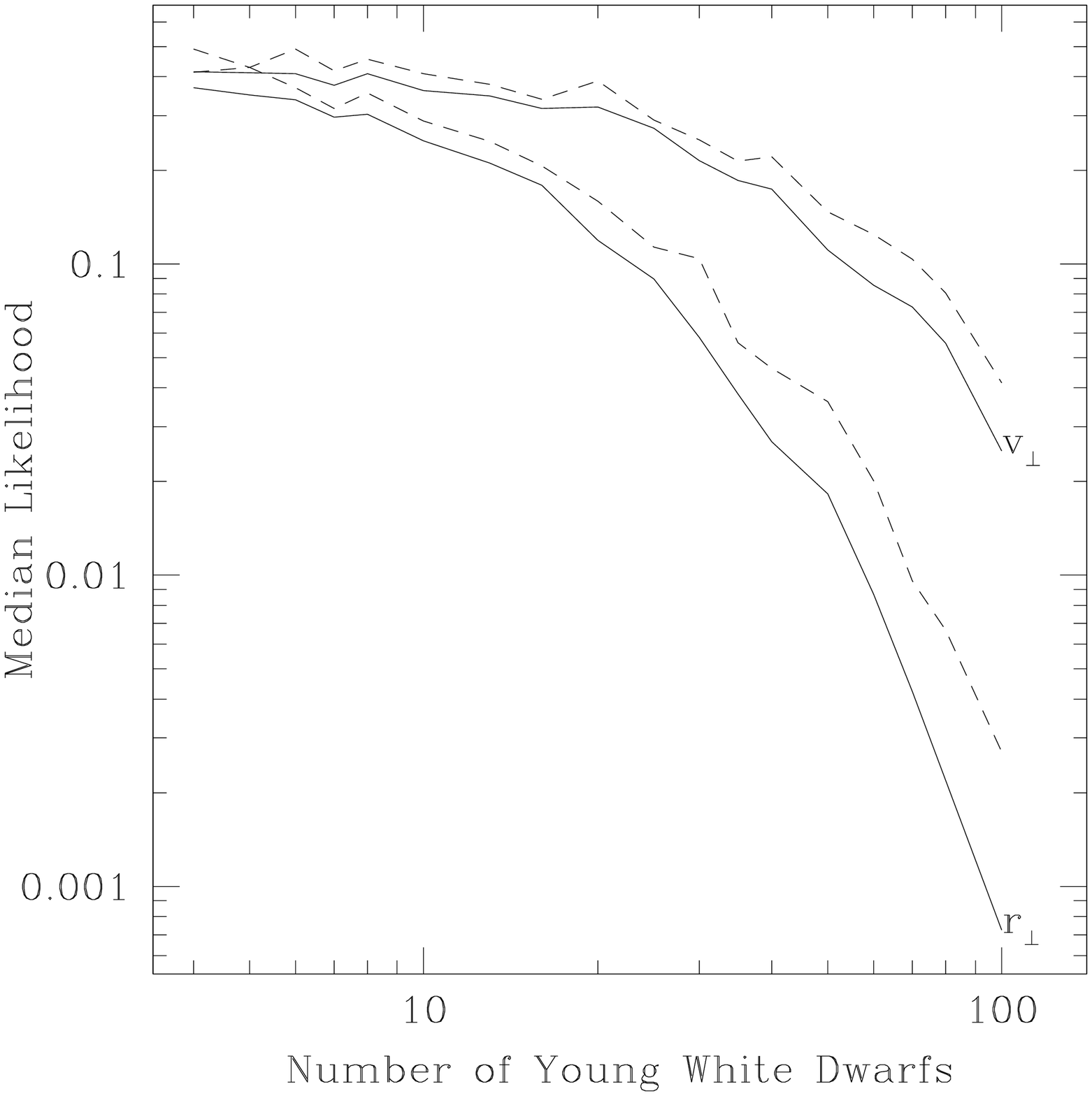}
\includegraphics[width=3.4in]{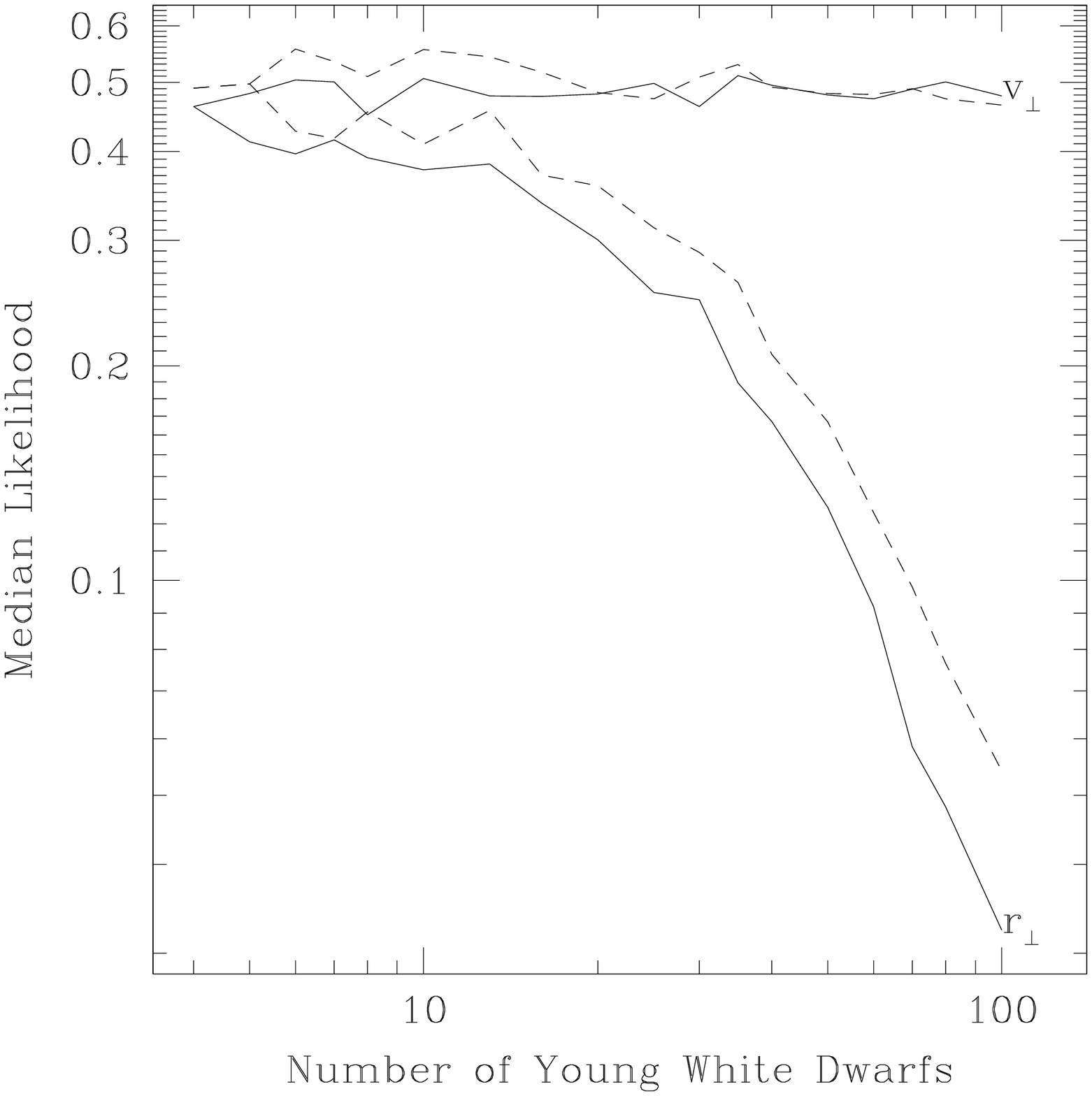}
\caption{The median significance of a detection of a difference
  between the distribution of young white dwarfs and main-sequence
  stars near the turn-off.  For the upper panel the initial
  distribution had $\sigma_\rmscr{TO}=0.5\sigma$ and a
  $\sigma_k=\sigma_\rmscr{TO}$.  For the lower panel the initial
  distribution had $\sigma_\rmscr{TO}=0.8\sigma$ and a
  $\sigma_k=\sigma_\rmscr{TO}$.  The cluster as a whole has
  $\Psi(0)=6\sigma^2$.  The dashed curves follow the median likelihood
  that the two realisations differ by chance (the probability of the
  null hypothesis) over 1,000 trials using the Kolmogorov-Smirnov test
  and the solid curves give the results using the Wilcoxon test.  The
  number of main-sequence stars in the sample is assumed to be ten
  times the number of white dwarfs.  }
\label{fig:significance_perp}
\end{figure}

The statistical likelihoods depicted in
Fig.~\ref{fig:significance_perp} of detecting the change in the
distributions support what is apparent from the distributions
themselves.  Finding a change in the radial distribution is much more
straightforward than finding changes in the velocity distribution
(even with perfect data).  Furthermore, for the loosely bound
progenitors (lower panel of Fig.~\ref{fig:significance_perp}), if it
is impossible to distinguish the proper motion distribution of the
young white dwarfs with kicks from their progenitors with fewer than
one white dwarfs in the sample (and 1,000 turn-off stars).  The key to
finding these changes is building up a sample of many stars.  For a
kick comparable to the initial velocity dispersion, at least one
hundred stars are required even in the best case for tightly bound
progenitor stars (upper panel of Fig.~\ref{fig:significance_perp}).

\subsection{Predictions}

\citet{2007Davis} reported a significant difference between the
distribution of young white dwarfs and turn-off stars and between
young white dwarfs and older white dwarfs.  Furthermore, they
estimated that the distribution of young white dwarfs was most similar
to that of main-sequence stars whose mass was half that of stars near
the turn-off.  \citet{Heyl07kickgc} found that such a large change in
the stellar distribution would require a kick with
$\sigma_k>\sigma_\rmscr{TO}$.  If one starts with
$\sigma_\rmscr{TO}=0.5\sigma$, a kick of $1.82\sigma_\rmscr{TO}$ is
required to make the final distribution of projected radii look like a
population with dispersion of $0.7\sigma$; that is one that
corresponds to one-half the mass of the progenitors.  The difference
between the initial and final distribution of radii is detectable with
a median likelihood of the null hypothesis of less than one part in
one thousand with only twenty young white dwarfs in the sample,
comparable to the probability quoted by \citet{2007Davis}.
\begin{figure}
\includegraphics[width=3.4in]{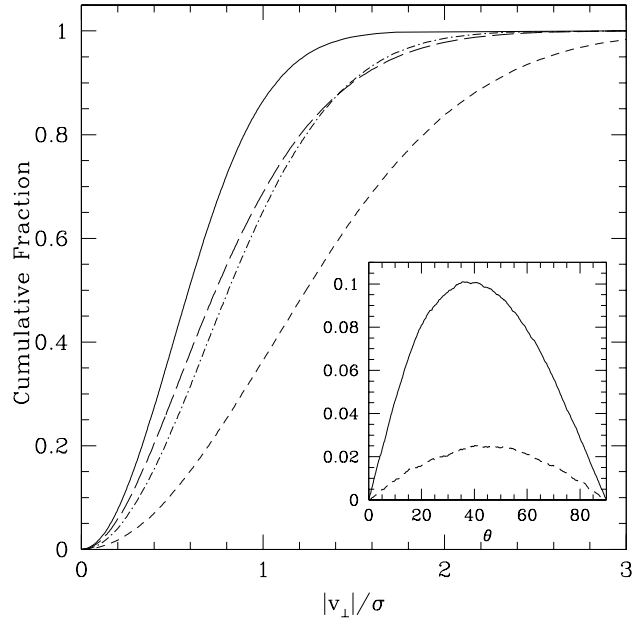}
\caption{The proper motion distribution for a white-dwarf kick large
  enough to account for the \citet{2007Davis} results
  ($\sigma_k=1.84\sigma_\rmscr{TO}$, $\sigma_\rmscr{TO}=0.5\sigma$ and
  $\Psi(0)=6\sigma^2$).  In the main figure (similar to the bottom
  panel of Fig.~\ref{fig:pos_vel}), the solid curves follow the
  initial distribution, the short-dashed curves follow the
  distribution immediately after the kick, and the long-dashed curves
  follow the distribution after phase mixing.  The dot-dashed curves
  follow the initial proper-motion distribution for a dispersion of
  $0.7\sigma$.  The inset shows the distribution of position angles of
  the proper motion (similar to Fig.~\ref{fig:angle}).  The solid
  curve gives the distribution if one neglects measurement errors and
  the dashed lines assumes a measurement error in the proper motion of
  $\sigma$ in each direction.  From Tab.~\ref{tab:gc} this translates
  to an accuracy of about one milliarcsecond per year for many nearby
  clusters.  }
  \label{fig:bigkick}
\end{figure}

With about forty young white dwarfs, the difference between the
initial and final proper motion distributions in
Fig.~\ref{fig:bigkick} is detectable with the likelihood of the null
hypothesis of less than two percent.  However, the systematics of
detecting an increase in the typical proper motion of the faint young
white dwarfs relative to the much brighter main sequence stars is
problematic because the expected errors in the astrometry increase as
the stars become fainter which is just the effect predicted for
white-dwarf kicks.  

On the other hand, an exploration of the direction of the proper motion
should not be affected by the systematics of errors in the astrometry
because the errors in the proper motion should be symmetric and even if
they are not symmetric it is unlikely that they will correlate with
the direction toward the centre of the cluster.  For the parameters
used here, about sixty percent of the stars have proper motions within
forty-five degrees of the radial direction, and about one hundred
young white dwarfs would be required to detect the asymmetry with the
median likelihood of the null hypothesis (Wilcoxon rank-sum text) of less
than two percent.  Although more white dwarfs are required for this
test, it is less plagued by systematics, so in practice it should be
more powerful.

In one includes a measurement error in the proper motion of $\sigma$
about and assumes that the error is isotropic, this necessarily
weakens the asymmetry.  On the other hand, these errors even if
poorly estimated cannot create a signal where there is none, in contrast
with measurements of the magnitude of the proper motion.  From
Tab.~\ref{tab:gc} this error translates to about one milliarcsecond
per year for nearby globular clusters, similar to the precision quoted
by \citet{2007arXiv0708.4030R}.  Including these realistic
errors one would expect about 52.5 percent of the stars to have proper
motions within forty-five degrees of the radial direction, so about
1,600 stars would be required to detect this asymmetry at the
two-sigma level if the astrometric errors in current instruments are
included.
\begin{table*}
  \caption{Kinetic parameters and distances for a few nearby globular clusters whose
    radial stellar distribution could be probed with JWST (Richer {\em
      priv. comm.}).  
    The concentration or the ratio of the core radius
    to the tidal radius ($c=\log_{10} r_t/r_c$) and the distance are from \citet{Harr96}, a
    dash denotes that the core has collapsed.
    The values of $\Psi(0)/\sigma_c^2$ are from
    \citet{2005ApJS..161..304M} with the exception of NGC 6837 where
    it was
    inferred from the concentration.   The value of
    $\sigma_c$ was obtained by assuming that ${\bar
      M}/M_\rmscr{TO}=0.3/0.8=0.375$. Refs: (a) \citet{2006ApJS..166..249M}; (b)
    \citet{2006A&A...445..503R}; (c) \citet{1986ApJ...305..645P}; (d) \citet{1993ASPC...50..357P}. }
\label{tab:gc}
\begin{center}
  \begin{tabular}{l|rrrrcc}
    \hline
\hline
\multicolumn{1}{c}{Cluster} &
\multicolumn{1}{c}{$\sigma_\rmscr{los}$ [km/s]} &
\multicolumn{1}{c}{$\sigma$  [km/s]} &
\multicolumn{1}{c}{$c$} &
\multicolumn{1}{c}{$\Psi(0)/\sigma_c^2$} &
\multicolumn{1}{c}{$d$ [kpc]} &
\multicolumn{1}{c}{$\sigma/d$ [milliarcseconds/yr]} \\
\hline 
NGC 104 (47 Tuc) & 11.6$\pm$0.8$^\rmscr{(a)}$ & 19 & 2.03 & 
8.60$\pm$0.10 & 4.5 & 0.89 \\
NGC 5139 (Omega Cen) & 15$^\rmscr{(b)}$ & 25 & 1.61 & 6.20$\pm$0.20 &
5.3 & 1.00 \\
NGC 6121 (M4) & 3.9$\pm$0.7$^\rmscr{(c)}$  & 6.4 & 1.59 &
7.40$\pm$0.10 & 2.2 & 0.61 \\
NGC 6397 & 3.5$\pm$0.2$^\rmscr{(d)}$ & 5.7 &  \multicolumn{1}{c}{---}
& \multicolumn{1}{c}{---} & 2.3 & 0.52 \\
NGC 6656 (M22) &  8.5$\pm$1.9$^\rmscr{(c)}$ & 14 & 1.31 &
6.50$\pm$0.20 & 3.2 & 0.92 \\
NGC 6752 & 4.5$\pm$0.5$^\rmscr{(d)}$ & 7.3 & \multicolumn{1}{c}{---}
&\multicolumn{1}{c}{---} & 4.0 & 0.38 \\
NGC 6809 (M55) & 4.2$\pm$0.5$^\rmscr{(d)}$ & 6.8 & 0.76 &
4.50$\pm$0.10 & 5.3 & 0.27 \\
NGC 6838 (M71)&  2.8$\pm$0.6$^\rmscr{(c)}$ & 4.6 &  1.15 & 5.4 & 4.0 & 0.18
\\
\hline
  \end{tabular}
\end{center}
\end{table*}

\section{Discussion}

If white dwarfs do indeed receive kicks as they form, both their
radial and velocity distributions will appear different from their
progenitors and the older white dwarfs whose distribution has relaxed.
The easier of these effects to measure is the change in the radial
distribution of the stars \citep{2007Davis}; however, there are two
other lines of evidence that might be present in astrometric
measurements of globular clusters.   First, white dwarfs typically
have larger proper motions than their progenitors and older white
dwarfs.  Unfortunately, the errors in the astrometry of young white
dwarfs are much larger than their much brighter progenitors on the main
sequence, so unless this systematic is handled carefully, this signal
is difficult to unravel.  Second, the white dwarfs after the kick are
more likely to lie on radial orbits than their progenitors.  Although
measurement errors will weaken observations of this effect, they
cannot mimic an asymmetry in the orbit parameters, so this signal
although weaker is less prone to systematics and may eventually
provide an independent line of evidence for white dwarf kicks.

Furthermore, these calculations indicate that we are on the verge of
being able to probe the dynamics of globular clusters in an entirely
new way through the direct measurement of the proper motions of the
stars.  Such measurements will provide unique information on the
structure of clusters as well as stellar evolution.

\section*{Acknowledgments}
I would like to thank Harvey Richer and Saul Davis for useful
discussions.  The Natural Sciences and Engineering Research Council of
Canada, Canadian Foundation for Innovation and the British Columbia
Knowledge Development Fund supported this work.  Correspondence and
requests for materials should be addressed to heyl@phas.ubc.ca.  This
research has made use of NASA's Astrophysics Data System Bibliographic
Services

\bibliographystyle{mn2e}
\bibliography{mine,wd,physics,math}
\label{lastpage}
\end{document}